\documentclass[12pt]{article}

\usepackage{amssymb}
\usepackage{amsmath}
\usepackage{amscd}
\usepackage{latexsym}
\usepackage{graphicx}

\usepackage{cite}
\newcommand{\be}{\begin{equation}}
\newcommand{\ee}{\end{equation}}

\newcommand{\prt}{\partial}

\newcommand{\bt}{\beta}
\newcommand{\vp}{\varphi}
\newcommand{\ep}{\varepsilon}
\newcommand{\al}{\alpha}
\newcommand{\ra}{\rightarrow}

\newcommand{\Gm}{\Gamma}

\begin{document}

\begin{center}
 
{\Large{\bf Strong-Coupling Extrapolation of Gell-Mann-Low Functions} \\ [5mm]

V.I. Yukalov$^{1,2}$ and E.P. Yukalova$^{3}$}  \\ [3mm]

{\it
$^1$Bogolubov Laboratory of Theoretical Physics, \\
Joint Institute for Nuclear Research, Dubna 141980, Russia \\ [2mm]

$^2$Instituto de Fisica de S\~ao Carlos, Universidade de S\~ao Paulo, \\
CP 369, S\~ao Carlos 13560-970, S\~ao Paulo, Brazil \\ [2mm]

$^3$Laboratory of Information Technologies, \\
Joint Institute for Nuclear Research, Dubna 141980, Russia } \\ [3mm]

{\bf E-mails}: {\it yukalov@theor.jinr.ru}, ~~ {\it yukalova@theor.jinr.ru} \\

\end{center}

\vskip 1cm

\begin{abstract}

Gell-Mann-Low functions can be calculated by means of perturbation theory and expressed 
as truncated series in powers of asymptotically small coupling parameters. However, 
it is necessary to know there behavior at finite values of the parameter and, moreover, 
their behavior at asymptotically large coupling parameters is also important. The problem 
of extrapolation of weak-coupling expansions to the region of finite and even infinite 
coupling parameters is considered. A method is suggested allowing for such an extrapolation. 
The basics of the method are described and illustrations of its applications are given 
for the examples where its accuracy and convergence can be checked. It is shown that in 
some cases the method allows for the exact reconstruction of the whole functions from 
their weak-coupling asymptotic expansions. Gell-Mann-Low functions in multicomponent 
field theory, quantum electrodynamics, and quantum chromodynamics are extrapolated to 
their strong-coupling limits.     
          
\end{abstract}

\newpage

\section{Introduction}

There exists a general problem, where, because of complexity, the sought solution can 
be calculated solely by means of perturbation theory resulting in truncated asymptotic 
series in powers of a coupling parameter or some variable. However, one needs to know 
the behavior of the solution at finite values of the parameter and, moreover, even for 
asymptotically large values of this parameter. There are several methods of extrapolation 
of asymptotic series to finite parameters but, it seems, there has been no reliable 
methods allowing for the extrapolation to infinite values of the parameters, when just 
a weak-coupling expansion is available and not so many terms of perturbation theory are 
known.

Consider, for instance, an expansion in powers of a variable $x \in [0,\infty)$. The 
popular Pad\'{e} approximation \cite{Baker_1} 
$$
P_{M/N}(x) ~= ~\frac{a_0+a_1 x+ a_2 x^2 +\ldots + a_M x^M}
{1+b_1 x+ b_2 x^2 +\ldots + b_N x^N}   
$$   
can provide reasonable accuracy for finite values of $x$, but for asymptotically large 
$x$, it behaves as
$$
P_{M/N}(x) ~= ~ \frac{a_M}{b_N} \; x^{M-N} \qquad ( x \ra \infty) \; .
$$   
Since $M$ and $N$ can be any integers, the large-variable behavior is not defined, being 
spread, depending on $M$ and $N$, between $-\infty$ and $\infty$. Such ambiguities are, 
actually, common for all methods of extrapolation, because of which the problem of finding 
the behavior of functions in the limit of $x \ra \infty$ is so much complicated. 

In the present communication, we describe a general method that makes it straightforward 
to extrapolate a weak-coupling expansion, containing just a few terms, to the whole range
of the variable, including the limit $x \ra \infty$. This approach is based on self-similar 
approximation theory \cite{Yukalov_2,Yukalov_3,Yukalov_4,Yukalov_5,Yukalov_38}. In the 
following section, we give a justification of the method we shall use.

\section{Method of Self-Similar Approximations}

Suppose, we are interested in the solution of a problem that can be treated only by means 
of perturbation theory with respect to an asymptotically small parameter or variable, 
resulting in an expansion
\be
\label{1}
f_k(x) ~= ~\sum_{n=0}^k a_n x^n \qquad ( x \ra 0 ) \; .
\ee
For concreteness, let us keep in mind a real function of a real variable $x\in[0,\infty)$.
Our goal is to extrapolate this asymptotic expansion to finite and even to infinite values
of the variable. As far as the region of large $x$ is the most difficult for treatment, at
the same time being often the most interesting, we shall concentrate on this limit.  
  
The basic idea is to consider the transition from one approximation order to another as
a motion of a dynamical system with respect to discrete time played by the approximation 
order. Then, knowing the law of motion, it could be possible to study the tendency of the
dynamical system for approaching an effective limit. The formation of such a dynamical 
system consists of the following steps.

(i) {\it Introduction of control functions}. The initially given sequence (\ref{1}), 
of course, cannot represent a stable dynamical system, since such sequences, as a rule, 
are divergent. Therefore the formation of a stable dynamical system has to start with 
the introduction of control parameters or control functions governing the sequence 
convergence. This process can be schematically denoted as a transformation 
\be
\label{2}
 F_k(x,u_k) ~= ~\hat T[\; u_k \;] \; f_k(x)  
\ee 
enjoying the inverse transformation
\be
\label{3}
 f_k(x) ~= ~\hat T^{-1}[\; u_k \;] \; F_k(x,u_k) \;  .
\ee
Control functions have to satisfy one of the two general conditions. Either they have 
to govern the transformed sequence convergence, or have to be defined from a training 
set of empirically known data. The first way is analogous to the introduction of 
controls for realizing controlled dynamical systems \cite{Bellman_6,Lee_7}. The second 
way is similar to the learning procedure in machine learning \cite{Murphy_8,Alpaydin_9}. 
The convergence condition is regulated by the Cauchy criterion saying that for a given 
$\varepsilon$ there exists $k_{\varepsilon}$ such that
\be
\label{4}
 |\; F_{k+p}(x,u_{k+p}) - F_k(x,u_k) \; | ~ < ~ \ep  
\ee
for any $k>k_{\varepsilon}$ and $p > 0$. The learning procedure assumes that, for 
a training set $\{z_k\}$, there are the known empirical data $\{F_k\}$, for which
\be
\label{5}
F_k(z_k,u_k) ~ = ~ F_k \;  .
\ee
   
(ii) {\it Definition of approximation endomorphisms}. This requires for the sequence 
of approximants $F_k(x,u_k)$ to put into correspondence a bijective sequence of 
endomorphisms $y_k(f)$ acting on the approximation space
\be
\label{6}
 {\cal A} = \overline {\cal L} \{ y_k(f) : ~ k = 0,1,2,\ldots \} \;  ,
\ee
which is a closed linear envelope of the approximation endomorphisms. The latter, by 
imposing a rheonomic constraint
\be
\label{7}
 f ~ = ~ F_0(x,u_k(x)) \; , \qquad x  ~ = ~ x_k(f) \; ,
\ee
are defined as
\be
\label{8}
 y_k(f)  ~ = ~ F_k(x_k(f),u_k(x_k(f) )) \;  .
\ee

(iii) {\it Formulation of evolution equation}. The Cauchy criterion (\ref{4}), in terms 
of the approximation endomorphisms (\ref{8}), reads as
\be
\label{9}
 |\; y_{k+p}(f) - y_k(f) \; |  ~ < ~  \ep \;.
\ee
The existence of a sequence limit implies that, with the increasing approximation order
$k$, one comes to a limiting value 
\be
\label{10}
y_{k+p}(f)  ~ \simeq ~ y^*(f) \; .
\ee
The sequence limit, for the approximation endomorphisms, is a fixed point, satisfying
the condition
\be
\label{11}
 y_k(y^*(f))  ~ = ~ y^*(f) \;  .
\ee
From these conditions, it follows that, in the vicinity of a fixed point, the self-similar
relation
\be
\label{12}
  y_{k+p}(f)  ~ = ~ y_k(y_p(f))  
\ee
is valid.      

In that way, we have a family of the approximation endomorphsims $y_k(f)$, acting on 
the approximation space (\ref{6}), with the approximation order $k$ playing the role 
of discrete time, and satisfying the self-similar evolution equation (\ref{12}). Such 
a dynamical system is called cascade \cite{Arrowsmith_10}, or in our case this is an 
{\it approximation cascade}
\be
\label{13}
\{ y_k(f) : ~\mathbb{Z}_+ \times {\cal A} \longmapsto {\cal A}\} \;   .
\ee
The points
\be
\label{14}
f \longmapsto y_1(f) \longmapsto y_2(f) \longmapsto \ldots
\longmapsto y_k(f) \longmapsto y_k^*(f)
\ee
form the cascade trajectory. Finding a fixed point of the cascade gives us the sequence 
limiting value $F_k^*(x,u_k)$. Performing the inverse transformation results in the 
self-similar approximant
\be
\label{15}
f_k^*(x) ~ = ~ \hat T^{-1}[\; u_k \;] \; F_k^*(x,u_k) \;  .
\ee
More details on the described techniques can be found in the recent reviews 
\cite{Yukalov_11,Yukalov_12}.

\section{Self-Similar Factor Approximants}

Control functions or control parameters can be introduced by means of different 
transformations. Presenting expansion (\ref{1}) in the form
\be
\label{16}
f_k(x) ~ = ~ \prod_{j=1}^k ( 1 + b_j x) \;   ,
\ee
we employ the fractal transform \cite{Barnsley_13}
\be
\label{17}
F_k(x,\{ n_j\} ) ~ = ~ \prod_{j=1}^k x^{-n_j} ( 1 + b_j x) \; .
\ee

Following the steps explained in the previous section, we come \cite{Yukalov_14,Gluzman_15}  
to the self-similar factor approximants 
\be
\label{18}
f_k^*(x) ~ = ~ \prod_{j=1}^{N_k} ( 1 + A_j x)^{n_j} \;  ,
\ee
where
\begin{eqnarray}
\label{19}
N_k ~ = ~ \left\{ \begin{array}{rr}
k/2 , ~ & ~ k = 2,4,\ldots \\
(k+1)/2 , ~ &  k = 3,5,\ldots 
\end{array} \right. \; .
\end{eqnarray}

Expression (\ref{18}) shows that, in order to define a real function $f_k^*(x)$, 
either $A_j$ and $n_j$ have to be real or, if they are complex valued, they need to 
enter the factor approximant in complex conjugate pairs, so that their product be real, 
taking into account the property
$$
| \; z^\al \; |^2 ~ = ~ | \; z \;|^{2{\Re}\al} 
\exp\{ - 2 ( {\Im}\al) {\rm arg} z \} \;  .
$$
Occasional complex approximants are to be discarded.
  
The parameters $A_j$ and $n_j$ are control parameters that can be defined from a training 
set played by the coefficients $a_n$ of the initial expansion (\ref{1}). The training 
conditions are
\be
\label{20}
  \lim_{x\ra 0} \; \frac{1}{n!} \frac{d^n}{dx^n} \; f_k^*(x) ~ = ~ a_n \; .
\ee
As is clear, these conditions guarantee the asymptotic equality
\be
\label{21}
f_k^*(x) ~ \simeq ~ f_k(x) \qquad ( x \ra 0 ) \; .
\ee
By comparing $\ln f_k^*(x)$ and $\ln f_k(x)$, the training conditions (\ref{20}) can be 
represented as the equations
\be
\label{22}
 \sum_{j=1}^{N_k} n_j A_j^m ~ = ~ L_m \qquad ( m = 1,2,\ldots,k ) \;  ,
\ee
where
\be
\label{23}
 L_m ~ = ~ 
\frac{(-1)^{m-1}}{(m-1)!} \; \lim_{x\ra 0} \; \frac{d^m}{dx^m} \; \ln f_k(x) \;  .
\ee

Being interested in the large-variable limit, we find from (\ref{18}) 
\be
\label{24}
f_k^*(x) ~ \simeq ~ B_k x^{\nu_k} \qquad ( x \ra \infty) \; ,
\ee
with the large-variable amplitude 
\be
\label{25}
B_k ~ = ~ \prod_{j=1}^{N_k} A_j^{n_j}
\ee
and the large-variable exponent
\be
\label{26}
\nu_k ~ = ~  \sum_{j=1}^{N_k} n_j \;  .
\ee
 
Equations (\ref{22}) uniquely define all control parameters for even orders $k=2,4,\ldots$. 
For odd orders of $k = 3,5,\ldots$, the number of equations $k$ is smaller than the number
$k+1$ of the parameters $A_j$ and $n_j$. To make the system of equations well defined, it 
is necessary to add one more equation. For this purpose, one can employ the diff-log
transformation of expansion (\ref{1}), construct the factor approximant for the transformed 
expansion, and to find the exponent $\nu_k$, thus getting one more equation \cite{Yukalov_16}. 
The convergence can be improved resorting to Borel transformation of series (\ref{1}).

\section{Zero-Dimensional Field Theory}

Before going to complicated problems, whose solutions are not known, let us first consider
the simpler cases, where it would be possible to check the convergence and accuracy of the
method. The popular test-horse is the so-called zero-dimensional $\varphi^4$ field theory,
with the generating functional
\be
\label{27}
 Z(g) ~ = ~ 
\frac{1}{\sqrt{\pi}} \int_{-\infty}^\infty \exp( -\vp^2 - g\vp^4)\; d\vp \;  ,
\ee
with the coupling $g \in [0,\infty)$. The weak-coupling expansion 
\be
\label{28}
Z_k(g) ~ = ~ \sum_{n=0}^k a_n g^n \qquad ( g \ra 0 )
\ee
strongly diverges for any finite coupling $g$, since the expansion coefficients 
\be
\label{29}
a_n ~ = ~ \frac{(-1)^n}{\sqrt{\pi}\; n!} \; \Gm(2n + 1)
\ee
factorially grow with $n$.   

We construct the self-similar factor approximants (\ref{18}) for expansion (\ref{28})
up to the $16$-th order and obtain the large-variable behavior of the functional 
$$
Z_{16}^*(g) ~ \simeq  ~ 0.828 g^{-0.187} \qquad ( g \ra \infty) \; ,
$$
whose accuracy is within about $20 \%$. Borel transformation improves convergence and 
accuracy, yielding in the $14$-th order
$$
Z_{14}^*(g) ~ \simeq  ~ 0.973 g^{-0.242} \qquad ( g \ra \infty) \;   ,
$$
whose error is around $3 \%$. This should be compared with the exactly known asymptotic 
behavior
$$
Z(g) ~ \simeq  ~ 1.023 g^{-0.25} \qquad ( g \ra \infty) \; .
$$
 
Although the accuracy of these results may seem to be not extremely impressive, however 
it is not as bad as well. In addition, one has to remember that no information has been 
used, except the bare weak-coupling expansion (\ref{28}). Moreover, calculations prove 
that even the low-order approximants give reasonable accuracy.

In any case, we have to accept that some hidden information is contained in the 
small-variable expansions of type (\ref{1}), and self-similar approximants do decode the 
hidden information, so that, even having nothing, except the small-variable expansion, 
the sought function can be restored for arbitrary variables, including the most difficult 
and important limit of asymptotically large variables.

\section{One-Dimensional Anharmonic Oscillator}

The other touch-stone for calculational methods is the one-dimensional anharmonic 
oscillator characterized by the Hamiltonian
\be
\label{30}
 H ~ =  ~ - \; \frac{1}{2} \; \frac{d^2}{dx^2} + \frac{1}{2} \; x^2 
+ g x^4  \;  ,
\ee
with $x\in(-\infty, \infty)$ and the coupling $g > 0$. The weak-coupling expansion of
the ground-state energy 
\be
\label{31}  
E_k(g)  ~ =  ~ \sum_{n=0}^k a_n g^n \qquad ( g \ra 0 )
\ee
strongly diverges because of the factorially growing coefficients that can be found in 
Ref. \cite{Bender_17}.

Directly applying self-similar factor approximants, we have in $14$-th order 
$$
E_{14}^*(g) ~ \simeq  ~ 0.739 g^{0.294} \qquad ( g \ra \infty) \;    ,
$$
within the error of $10\%$. Involving Borel transformation, in the same order, we get
$$
E_{14}^*(g) ~ \simeq  ~ 0.688 g^{0.327} \qquad ( g \ra \infty) \;   ,   
$$ 
with the error of about $2\%$. The exact asymptotic behavior is
$$
E(g) ~ \simeq  ~ 0.668 g^{1/3} \qquad ( g \ra \infty)  \;  .
$$

Again, being based solely on the weak-coupling expansion, self-similar approximants restore
the sought function for all couplings, including asymptotically large.

\section{Supersymmetric Yang-Mills Theory}

In some cases, self-similar approximants can restore the whole function exactly, 
provided we can compare the obtained approximants with the known exact expression 
\cite{Yukalov_39,Yukalov_18}. Let us consider the $N = 1$ symmetric pure Yang-Mills 
theory, where the weak-coupling expansion for the Gell-Mann-Low function reads as 
\cite{Novikov_19,Shifman_20,Arkani_21,Arkani_22}
\be
\label{32}
 \bt_k(g) ~ = ~ - \; \frac{3g^3 N_c}{16\pi^2} \sum_{n=0}^k b_n g^{2n} \qquad
( g \ra 0 ) \;  ,
\ee
with the coefficients $b_n = (N_c/8 \pi^2)^n$. 

The self-similar approximant of second order is
\be
\label{33}
 \bt_2^*(g) ~ = ~ - \; 
\frac{3g^3 N_c}{16\pi^2} \; \left( 1 + A_1 g^2 \right)^{n_1} \;   .
\ee
From the training set of the given $b_n$, we have the control functions
$$
A_1 ~ = ~ - \; \frac{N_c}{8\pi^2} \; , \qquad n_1 ~ = ~ - 1\; .
$$
Thus the self-similar factor approximant takes the form
\be
\label{34}
 \bt_2^*(g) ~ = ~ - \; 
\frac{3g^3 N_c}{16\pi^2} \; 
\left( 1 - \; \frac{N_c}{8\pi^2}\; g^2 \right)^{-1} \;   ,
\ee
which coincides with the exact known beta function 
\cite{Novikov_19,Shifman_20,Arkani_21,Arkani_22}. All higher orders of the approximants
are also reduced to the exact expression (\ref{34}),
\be
\label{35}
 \bt_k^*(g) ~ = ~ \bt_2^*(g) \qquad ( k = 2,3,\ldots ) \; .
\ee

\section{Multicomponent Field Theory}

The weak-coupling expansion of the Gell-Mann-Low function in the $N$-component $\phi^4$ 
theory, is 
\be
\label{36}
 \bt_k(g) ~ = ~ g^2 \sum_{n=0}^k b_n g^n \qquad ( g \ra 0 ) \; ,
\ee
with the coefficients within a minimal subtraction scheme, in the six-loop approximation,
given in Ref. \cite{Kompaniets_23}. For $N = 1$, the Gell-Mann-Low function is known in 
the seven-loop approximation \cite{Schnetz_24}. 

The self-similar factor approximants, for $N=1,2,3,4$, in the strong-coupling limit yield
$$
\bt_5^*(g) ~ \simeq ~ 1.698 g^{1.764} \qquad ( N = 0 ) \; ,
$$
$$
\bt_6^*(g) ~ \simeq ~ 1.857 g^{1.750} \qquad ( N = 1 ) \; ,
$$
$$
\bt_5^*(g) ~ \simeq ~ 2.017 g^{1.735} \qquad ( N = 2 ) \; ,
$$
$$
\bt_5^*(g) ~ \simeq ~ 2.178 g^{1.719} \qquad ( N = 3 ) \; ,
$$
\be
\label{37}
\bt_5^*(g) ~ \simeq ~ 2.340 g^{1.702} \qquad ( N = 4 ) \;   .
\ee
Using the self-similar Borel summation gives
$$
\bt_4^*(g) ~ \simeq ~ 1.42 g^{1.83} \qquad ( N = 0 ) \; ,
$$
$$
\bt_6^*(g) ~ \simeq ~ 1.70 g^{1.77} \qquad ( N = 1 ) \; ,
$$
$$
\bt_5^*(g) ~ \simeq ~ 2.18 g^{1.67} \qquad ( N = 2 ) \; ,
$$
$$
\bt_5^*(g) ~ \simeq ~ 2.50 g^{1.63} \qquad ( N = 3 ) \; ,
$$
\be
\label{38}
\bt_5^*(g) ~ \simeq ~ 2.81 g^{1.61} \qquad ( N = 4 ) \; .
\ee

No exact results are known for the strong-coupling behavior of this function. In that 
sense, the behavior of the beta functions at large $g \ra \infty$ is a prediction. This 
prediction uses the self-similar factor approximant that is trained on the data from 
the weak-coupling region. This procedure is somewhat similar to the forecasting made 
by means of dynamical models trained on a set of initial data 
\cite{Montgomery_25,Makridakis_26,Ascher_27} and to machine-learning trained on these 
data \cite{Murphy_8,Alpaydin_9}. 

Some estimates for the value of the coupling $g \gtrsim 1$ have been done for $N=1$ in 
Refs. \cite{Kazakov_28,Chetyrkin_29}. They used a Borel-type summation with conformal 
mapping, including three control parameters chosen so that to make the Borel transform
self-consistent and satisfying the minimal-difference condition. The latter is known 
to improve convergence, being a part of optimized perturbation theory 
\cite{Yukalov_30,Yukalov_31}. The behavior of the beta function for $g \gtrsim 1$ was 
found to be proportional to $g^\nu$, with $\nu \in [1.7, 2.2]$.       

It is worth stressing that in the method of self-similar approximants not merely the 
strong-coupling limit can be evaluated, as in Refs. \cite{Kazakov_28,Chetyrkin_29}, but 
the whole function is obtained. For example, in the case of $N = 1$, in the second and 
sixth order we have
$$
\bt_2^*(g) ~ = ~ \frac{3g^2}{(1+9.599 g)^{0.197}} \; , 
$$
\be
\label{39}
\bt_6^*(g) ~ = ~ \frac{3g^2}{(1+5.357 g)^{0.187}(1+13.72 g)^{0.06}(1+22.096 g)^{0.003}} 
 \qquad  ( N = 1 ) \; .
\ee

From the renormalization group equation
\be
\label{40}
\mu \; \frac{\prt g}{\prt\mu} ~ = ~ \bt(g) \;  ,
\ee
where $\mu$ is a scale parameter, it follows that, under $\nu > 1$, when the coupling 
rises, so that
\be
\label{41}
\bt(g) ~ \simeq ~ B g^\nu \qquad ( g \ra \infty) \;  ,
\ee
then $g$ increases as  
\be
\label{42}
g ~ \simeq ~ \frac{1}{[(\nu-1)B\ln(\mu_0/\mu)]^{1/(\nu-1)}} \qquad
( \mu \ra \mu_0 - 0 ) \; .
\ee

\section{Quantum Electrodynamics}

The Gell-Mann-Low function in quantum electrodynamics, in the renormalized minimal 
subtraction scheme, has the weak-coupling five-loop expansion
\be
\label{43}
\bt_k(\al) ~ = ~ \left( \frac{\al}{\pi}\right)^2
\sum_{n=0}^k b_n \left( \frac{\al}{\pi}\right)^n \;   .
\ee
One usually takes into account electrons, although the contributions of leptons with 
higher masses, such as muons and tau leptons, are neglected. The coefficients $b_n$ are 
given in Ref. \cite{Kataev_32}.

Self-similar factor approximants lead to the strong-coupling limit
\be
\label{44}
 \bt_4^*(\al) ~ \simeq ~  0.476 \left( \frac{\al}{\pi}\right)^{2.096} 
\qquad \left( \frac{\al}{\pi} \ra \infty \right) \; ,
\ee
and the self-similar Borel summation results in the strong-coupling behavior
\be
\label{45}
 \bt_3^*(\al) ~ \simeq ~  0.587 \left( \frac{\al}{\pi}\right)^{2.12} 
\qquad \left( \frac{\al}{\pi} \ra \infty \right) \;   .
\ee

The QED running coupling is the solution to the renormalization group equation
\be
\label{46}
 \mu^2 \; \frac{\prt}{\prt\mu^2} \left( \frac{\al}{\pi}\right) ~ = ~  
\bt(\al) \;  .
\ee
As an initial condition, it is possible to take $\alpha$ at the Z-boson mass,
$$
 \al(m_Z) ~ = ~  0.007815 \; , \qquad m_Z ~ =  ~  91.1876 \; {\rm GeV} \;  .
$$
Then the behavior of the coupling in the vicinity of the scale point
\be
\label{47}
\mu_0 ~ =  ~  8.58 \times 10^{260} \; {\rm GeV} \;   ,
\ee
where the beta function sharply increases as
\be
\label{48}
\bt(\al) ~ =  ~ B \left( \frac{\al}{\pi}\right)^\nu \qquad 
\left( \frac{\al}{\pi} \ra \infty \right) \;  ,
\ee
can be written in the form
\be
\label{49}
\al ~ \simeq ~  \frac{\pi}{[(\nu-1)2B\ln(\mu_0/\mu)]^{1/(\nu-1)}} \; .
\ee
Substituting here numerical values gives
\be
\label{50}
\al ~ \simeq ~ \frac{2.743}{[\ln(\mu_0/\mu)]^{0.682}} \qquad
(\mu \ra \mu_0 - 0 ) \; .
\ee
As is seen, the point of divergence (\ref{47}) is drastically shifted from the Landau
pole that is of order $10^{30}$ Gev \cite{Deur_33}.

\section{Quantum Chromodynamics}

The weak-coupling expansion of the Gell-Mann-Low function in quantum chromodynamics
reads as
\be
\label{51}
\bt_k(\al_s) ~ = ~ - \; \left( \frac{\al_s}{\pi}\right)^2
\sum_{n=0}^k b_n \left( \frac{\al_s}{\pi}\right)^n 
\qquad
\left( \frac{\al_s}{\pi} \ra 0 \right) \;   ,
\ee
where $\alpha_s$ is the quark-gluon coupling. The coefficients, within the minimal 
subtraction scheme, can be found in Refs. \cite{Luthe_34,Baikov_35,Herzog_36}. We keep 
in mind the physically realistic case of $N_c =3$ colors and $n_f = 6$ flavors. Bound
states are not considered.

Self-similar approximation results in the strong-coupling behavior of the beta function
\be
\label{52}
\bt_2^*(\al_s) ~ \simeq ~  - 2.277 \al_s^{2.598} \qquad 
\left( \frac{\al_s}{\pi} \ra \infty \right) \;   .
\ee
Self-similar Borel summation leads to
\be
\label{53}
\bt_2^*(\al_s) ~ \simeq ~  - 1.89 \al_s^{2.75} \qquad 
\left( \frac{\al_s}{\pi} \ra \infty \right) \;    .
\ee

The QCD running coupling satisfies the renormalization group equation 
\be
\label{54}
 \mu^2 \; \frac{\prt}{\prt\mu^2} \left( \frac{\al_s}{\pi}\right) ~ = ~  
\bt(\al_s) \;   .
\ee
As an initial condition, one can take the value of the coupling at the mass of the Z-boson,
$$
  \al_s(m_Z) ~ = ~  0.1184 \; , \qquad m_Z ~ =  ~  91.1876 \; {\rm GeV} \;  .
$$
Then the running coupling in the vicinity of $\mu_c = 0.1$ GeV behaves as
\be
\label{55}
\al_s ~ \simeq ~ \frac{0.907}{[\ln(\mu/\mu_c)]^{0.626}} \qquad
(\mu \ra \mu_c + 0 ) \; .
\ee
Again, the point of divergence is essentially shifted from the Landau pole that is close
to $0.9$ GeV \cite{Tanabashi_37}.

\section{Conclusion}

Self-similar approximation theory is presented allowing for the extrapolation of functions
from weak-coupling expansions to the whole range of their variables, including the limit
of asymptotically large variables tending to infinity. The justification of the approach
is given. Briefly speaking, the approach is based on discovering self-similarity in the
coefficients of the considered expansion, which then allows for its effective extrapolation  
to higher orders. 

The use of self-similar factor approximants is demonstrated for the problems, whose exact 
behavior is known. It is shown that in some cases, e.g. in the case of supersymmetric 
Yang-Mills theory, the beta function can be reconstructed exactly. The strong-coupling 
extrapolation for the Gell-Mann-Low functions is accomplished for the $N$-component 
$\varphi^4$ theory, quantum electrodynamics, and quantum chromodynamics.    

\vskip 5mm

{\bf Conflict of interest}

\vskip 2mm
 
The authors declare that they have no conflicts of interest.

\end{document}